\documentclass[traditabstract]{aa}
\usepackage{graphicx}
\usepackage{txfonts}

 \newcommand{\ngc}{NGC\,1433}

\def\tex {\ifmmode{{T}_{\rm ex}}\else{$T_{\rm ex}$}\fi}
\def\tmb {\ifmmode{{T}_{\rm mb}}\else{$T_{\rm mb}$}\fi}
\def\ci     {\ifmmode{{\rm C}{\rm \small I}}\else{C\ts {\scriptsize I}}\fi}
\def\hi     {\ifmmode{{\rm H}{\rm \small I}}\else{H\ts {\scriptsize I}}\fi}
\def\hh     {\ifmmode{{\rm H}_2}\else{H$_2$}\fi}

\def\ts     {\thinspace}
\def\kms    {\ifmmode{{\rm \ts km\ts s}^{-1}}\else{\ts km\ts s$^{-1}$}\fi}
\def\msol   {\ifmmode{{\rm M}_{\odot}}\else{M$_{\odot}$}\fi}
\def\lsol   {\ifmmode{{\rm L}_{\odot}}\else{L$_{\odot}$}\fi}
\def\zsol   {\ifmmode{{\rm Z}_{\odot}}\else{Z$_{\odot}$}\fi}
\def\etal   {{\rm et\ts al.}~}

\begin{document}

\title{ALMA observations of feeding and feedback in nearby Seyfert galaxies: an AGN-driven outflow in \ngc\
\thanks{Based on observations carried out with  ALMA in Cycle 0.
}}

\author{F. Combes \inst{1}
\and
S. Garc\'{\i}a-Burillo \inst{2}
\and
V. Casasola \inst{3}
\and
L. Hunt \inst{4}
\and
M. Krips \inst{5}
\and
A. J. Baker \inst{6}
\and
F. Boone \inst{7}
\and
A. Eckart \inst{8}
\and
I. Marquez \inst{9}
\and
R. Neri \inst{5}
\and
E. Schinnerer \inst{10}
\and
L. J. Tacconi \inst{11}
           }
\offprints{F. Combes}
\institute{Observatoire de Paris, LERMA (CNRS:UMR8112), 61 Av. de l'Observatoire, F-75014, Paris, France
\email{francoise.combes@obspm.fr}
 \and
Observatorio Astron\'omico Nacional (OAN)-Observatorio de Madrid,
Alfonso XII, 3, 28014-Madrid, Spain
 \and
INAF -- Istituto di Radioastronomia \& Italian ALMA Regional Centre, via Gobetti 101, 40129, Bologna, Italy
 \and
INAF - Osservatorio Astrofisico di Arcetri, Largo E. Fermi, 5, 50125, Firenze, Italy
 \and
IRAM, 300 rue de la Piscine, Domaine Universitaire, F-38406 Saint Martin d'H\`eres, France
 \and
Dep. of Physics \& Astronomy, Rutgers, the State University of New Jersey, 136 Frelinghuysen road, Piscataway, NJ 08854, USA
 \and
CNRS, IRAP, 9 Av. colonel Roche, BP 44346, 31028, Toulouse Cedex 4, France
 \and
I. Physikalisches Institut, Universit\"at zu K\"oln, Z\"ulpicher Str. 77, 50937, K\"oln, Germany
 \and
Instituto de Astrofisica de Andaluc{\'{\i}}a (CSIC), Apdo 3004, 18080 Granada, Spain
 \and
Max-Planck-Institut f\"ur Astronomie (MPIA), K\"onigstuhl 17, 69117 Heidelberg, Germany
 \and
Max-Planck-Institut f\"ur extraterrestrische Physik, Giessenbachstr. 1, Garching bei M\"unchen, Germany
              }

   \date{Received  2013/ Accepted  2013}

   \titlerunning{CO in NGC~1433}
   \authorrunning{F. Combes et al.}

   \abstract{We report ALMA observations of CO(3-2) emission in the Seyfert 2 galaxy \ngc\
at the unprecedented spatial resolution of 0\farcs5\,=\,24~pc. 
Our aim is to probe AGN (active galactic nucleus) feeding and feedback phenomena
through the morphology and
dynamics of the gas inside the central kpc.  The galaxy
\ngc\ is a strongly barred spiral with three resonant rings: one at the ultra-harmonic resonance 
near corotation, and the others at the outer and inner
Lindblad resonances (OLR and ILR). A nuclear bar of 400~pc radius is embedded in the large-scale primary bar.
The CO map, which covers the whole nuclear region (nuclear bar and ring), 
reveals a nuclear gaseous spiral structure, inside the nuclear ring encircling the nuclear stellar bar. 
 This gaseous spiral is well correlated with the dusty spiral seen in Hubble Space Telescope images. The nuclear spiral
winds up in a pseudo-ring at $\sim$200~pc radius, which might correspond to the inner ILR. Continuum emission is
detected at 0.87~mm only at the very centre, and its origin is more likely  
thermal dust emission than non-thermal emission from the AGN. It might correspond to the molecular torus
expected to exist in this Seyfert 2 galaxy.
The HCN(4-3) and HCO$^+$(4-3) lines were observed simultaneously,
but only upper limits are derived, with a ratio to the CO(3-2) line lower than 1/60 at 3$\sigma$,
indicating a relatively low abundance of very dense gas.
The kinematics of the gas over the nuclear disk reveal rather regular rotation only slightly perturbed by streaming
motions due to the spiral; the primary and secondary bars are too closely aligned with the galaxy
major or minor axis to leave a signature in the projected velocities. Near the nucleus, there is an intense high-velocity 
CO emission feature redshifted to 200~km/s
(if located in the plane), with a blue-shifted counterpart, at 2\arcsec\ (100~pc) from the centre. 
While the CO spectra
are quite narrow in the centre, this wide component is interpreted as an outflow involving a molecular mass
of 3.6 10$^6$ M$_\odot$ and a flow rate $\sim$ 7 M$_\odot$/yr. The flow could be in part driven by
the central star formation, but is mainly boosted by the AGN through its wind or radio jets.
}
\maketitle


\section{Introduction}

It is now observationally well established that supermassive black holes (SMBHs) reside in the nuclei of all 
galaxies with massive spheroids in the Local Universe and at higher redshifts as well (e.g. Kormendy \& Ho 2013).
Quasars at high redshift and 
Seyfert nuclei locally are fueled by accretion of material onto the SMBH. Although much progress has been made on 
both theoretical and observational fronts in the last decade, the relationship of black hole growth with galaxy 
formation and evolution is still far from being completely understood. 

One of the outstanding problems is to identify the mechanism that drives gas from the disk towards the nucleus, 
removing its large angular momentum,  to feed the central black hole and trigger the 
nuclear activity. Theoretically, broad-brush solutions have been found; cosmological simulations rely on merger-driven 
gas inflow via bar instabilities to feed a central starburst and fuel the SMBH 
(e.g., Hopkins \etal  2006; di Matteo \etal 2008). Nevertheless, in the Local Universe, no 
clear correlation has been found between the presence of
an active galactic nucleus (AGN) and either companions or the presence of bars (see e.g. Combes 2003, 2006, 
Jogee 2006 for reviews). It is possible that locally the relation between these large-scale phenomena and the 
duty cycle of nuclear fueling is masked by different timescales. Indeed, the presence of resonant rings, 
vestiges of a previous bar, appears to be correlated with Seyfert activity (Hunt \& Malkan 1999). It could  also
be that gas inflow is not always possible because of dynamical barriers (e.g., nuclear rings, see 
Piner \etal 1995; Regan \& Teuben 2004).

To assess potential inhibitors of the ubiquitous gas inflow assumed in simulations, we must examine the 
nuclear kinematics around local AGN. This can be best done with molecular tracers, since in galaxy centres 
HI is typically converted to molecular gas;  CO line emission is therefore our best probe, 
and in particular CO(3-2), which traces the high density gas (10$^4$-10$^5$ cm$^{-3}$) in the dense AGN 
circumnuclear regions (as we have shown in Boone \etal 2011). The HCN and HCO$^+$ 
line emission should trace the densest material  (at least 10$^7$ cm$^{-3}$),
and diagnose its excitation and chemistry. We have undertaken during the last 
decade the NUGA (NUclei of GAlaxies) program to study the gas distributions in nearby AGN, and find clues to 
their fueling. In the dozen nearby Seyfert or LINER galaxies observed with the IRAM Plateau de Bure 
interferometer (PdBI) in 
CO(2-1), we achieved a spatial resolution of 50-100~pc, and frequently worse 
for the most distant galaxies. In these galaxies, a large variety of gas distributions have 
been found; however, we detected on-going AGN feeding at 0.1-1~kpc scales for only five out of twelve cases: 
NGC\,2782 (Hunt \etal 2008, bar triggered by an interaction), 
NGC\,3147 (Casasola \etal 2008), 
NGC\,3627 (Casasola \etal 2011), 
NGC\,4579 (Garc\`{\i}a-Burillo \etal 2009), 
and 
NGC\,6574 (Lindt-Krieg \etal 2008).
The most common feeding mechanism 
in these galaxies appears to be kinematically decoupled embedded bars, i.e.  the combination of a slowly 
rotating kpc-scale stellar bar (or oval)  and a kinematically decoupled nuclear bar,
with overlapping dynamical resonances. Such resonances and kinematic decoupling are fostered by a large 
central mass concentration 
and high gas fraction. The gas is first stalled in a nuclear ring (a few 100~pc scale), and then driven 
inward under the influence of the decoupled nuclear bar. 
However, because of insufficient resolution,
our previous observations were unable
to probe the gas within 100~pc of the AGN most of the time.

In this paper, we present ALMA Cycle 0 observations in the CO(3-2) line of the Seyfert 2 \ngc, where the beam
is 24~pc in size.
The nearby distance (9.9~Mpc) and low inclination of 33$^\circ$ make \ngc\ an ideal target to test and 
refine the scenario of AGN feeding and feedback, and to constrain BH models which are only now beginning to examine 
in detail gas structures within 100 pc (Hopkins \& Quataert 2010; Perez-Beaupuits \etal 2011). 

Up to now, resolution at scales of tens of pc has been obtained only in a few Seyfert galaxies, and only 
in hot or warm gas tracers. 
The best example is NGC\,1068, the closest Seyfert 2 prototype, where near-IR 
\hh\ lines have been mapped with SINFONI at 0\farcs075 resolution (5.2pc) by M\"uller-Sanchez \etal (2009). 
The behaviour of this hot (1000-2000K) gas is not yet settled, however; while an outflow model is proposed by 
Galliano \& Alloin (2002), and a warped disk model by Schinnerer \etal (2000), M\"uller-Sanchez \etal (2009) propose 
a strong inflow model. Krips \etal (2011) explain their SMA CO(3-2) map at  0\farcs6\,=\,40~pc resolution by a 
rotating disk plus an outflow of the disk gas due to shocks and/or a circumnuclear disk-jet interaction.  
Thus, gas inflow could fuel the AGN at a 10-pc scale, and the jet-gas interaction could simultaneously drag gas outwards 
on scales of hundreds of pc.
It has also been suggested that the
 presence of an outflow in the circumnuclear disk of NGC\,1068 is 
responsible for the large-scale molecular shocks revealed by 
strong SiO emission in this galaxy (Garc\`{\i}a-Burillo \etal 2010).
The outflow is clearly seen in ALMA data (Garc\`{\i}a-Burillo \etal 2013, in prep.).
Typical outflow velocities are found of the order of 200~km/s in NGC\,1068.

An important ingredient in cosmological simulations is feedback, which can regulate SMBH growth and 
suppress star formation (e.g., Croton \etal 2006, di Matteo \etal 2008, and references therein). 
Molecular observations can constrain specific feedback mechanisms by discovering molecular outflows 
through their high-velocity wings, and can determine their origin (star formation or AGN) through high resolution observations. 
Chung \etal (2011) showed the ubiquitous presence of 1000~km/s molecular outflows in starbursts with SFRs larger 
than 100~M$_\odot$/yr (see also Feruglio \etal 2010, Fischer \etal 2010, Sturm \etal 2011).
The CO emission in the high-velocity wings may generally represent 25\% 
of the total observed emission. In NGC 1068 the outflow, if present, is only of the order of 200~km/s and is entrained by 
the radio jet. Coil \etal (2011) also find that galactic winds are frequent in ionized gas lines in post-starburst 
and AGN host galaxies at 0.2$<$z$<$0.8, but they are low-velocity winds, likely due to supernovae. 
High-velocity winds, driven by an AGN, might be frequent in molecular gas (Leon \etal 2007, Feruglio \etal 2010, 2013, 
Alatalo \etal 2011, Nesvadba \etal 2011, Dasyra \& Combes 2012, Aalto \etal 2012,
Spoon \etal 2013, Veilleux \etal 2013), 
as well as in the ionized or atomic gas component (Rupke \etal 2005; Riffel \& Storchi-Bergmann 2011). 
In Arp 220, Sakamoto \etal (2009) have discovered 100-200~km/s 
outflows through P-Cygni profiles in HCO$^+$(3-2), HCO$^+$(4-3), and CO(3-2) along the line of sight to the nucleus. 
They interpret this gas as driven outwards by the nuclear starburst.
Because \ngc\ is not an IR-luminous starburst, 
it is unlikely that
an AGN wind close to the nucleus would be 
swamped by a starburst wind, 
thus facilitating its identification with ALMA's high resolution.

\begin{figure*}[ht]
\centerline{
\includegraphics[angle=-90,width=15cm]{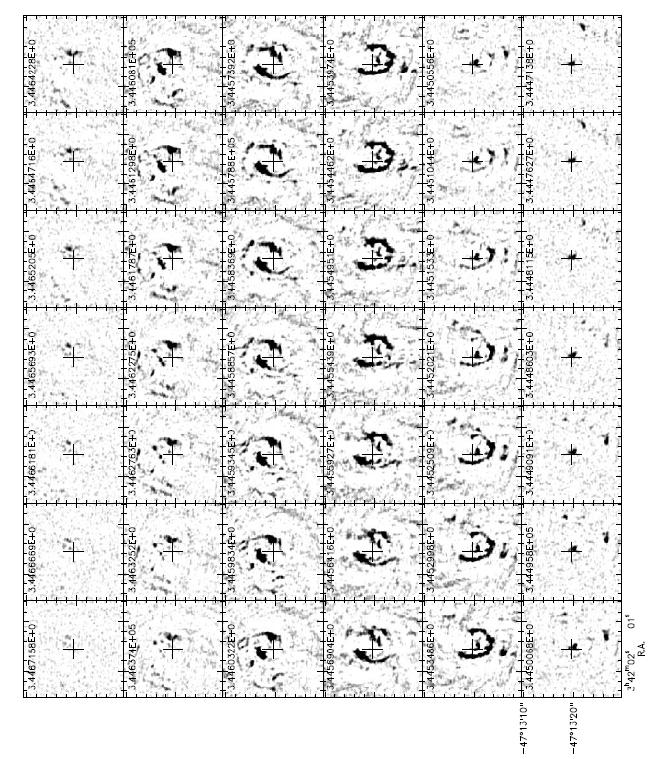}}
\caption{ Channel maps of CO(3-2) emission in the centre of \ngc.
Each of the 42 square boxes is 20\arcsec\, in size, while the primary beam
is 18\arcsec\, in diameter . Channels are separated
by 4.24~km/s. They are plotted from 978 (top left) to 1152~km/s (bottom right, 
the panels are labelled in frequency).
The synthesized beam is 0\farcs56$\times$0\farcs42 (PA=85$^\circ$). The centre of the maps 
is the phase centre of the interferometric observations 
given in Table \ref{tab:basic}. The colour scale is linear, between 1 and 30 mJy/beam.
}
\label{fig:chann}
\end{figure*}

\begin{figure}[ht]
\centerline{
\includegraphics[angle=-0,width=8cm]{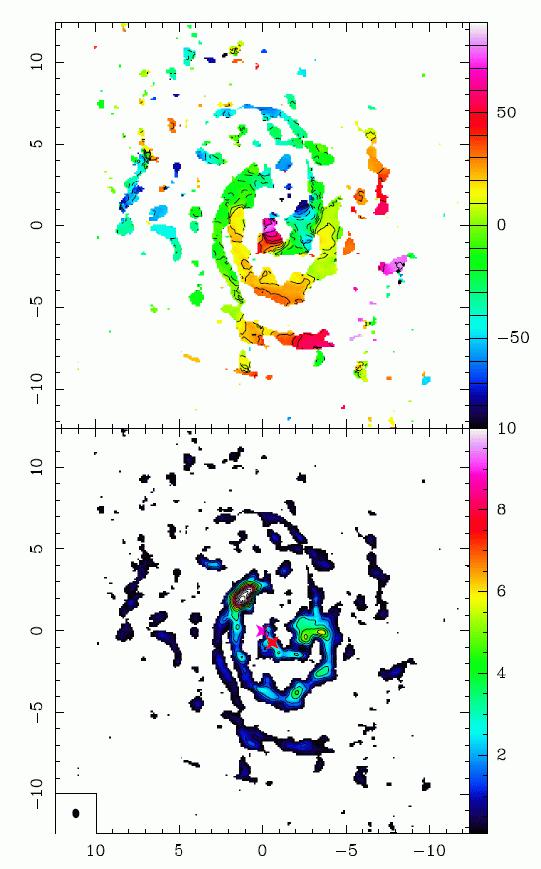}}
\caption{ Velocity field (top) and integrated intensity (bottom) of the CO(3-2) emission in 
the centre of \ngc. Coordinates are RA-Dec in arcsec relative to the phase centre 
(see Table \ref{tab:basic}). The colour scale ranges are in km/s relative to 1075~km/s (top) and 
in Jy/beam.MHz (or 0.87 Jy/beam.km/s) at the bottom. The beam size of 
0\farcs56$\times$0\farcs42 is indicated at the bottom left. The phase centre is 
a pink cross, and the new adopted centre is the red cross, shown in the bottom panel.
}
\label{fig:velo}
\end{figure}

%
\begin{center}
\begin{table}
      \caption[]{Basic data for the \ngc\ galaxy} 
\label{tab:basic}
\begin{tabular}{lll}
\hline
Parameter  & Value$^{\mathrm{b}}$ & Reference$^{\mathrm{c}}$ \\
\hline
$\alpha_{\rm J2000}$$^{\mathrm{a}}$ & 03$^h$42$^m$01.55$^s$ & (1) \\
$\delta_{\rm J2000}$$^{\mathrm{a}}$ &-47$^{\circ}$13$^{\prime}$19.5\arcsec\ & (1) \\
$V_{\rm hel}$ & 1075 km\,s$^{-1}$ & (1) \\
RC3 Type & (R')SB(r)ab & (1) \\
Nuclear Activity & Seyfert 2 & (2) \\
Inclination & 33\fdg0 & (3) \\
Position Angle & 199$^{\circ}$ $\pm$ 1$^{\circ}$ & (3) \\
Distance & 9.9\,Mpc (1\arcsec\ = 48\,pc) & (4) \\
L$_{B}$        & $1.0 \times 10^{10}$\,L$_{\odot}$ & (4) \\
M$_{\rm H\,I}$ & $5.5 \times 10^{8}$\,M$_{\odot}$ & (5) \\
M$_{\rm H_{2}}$& $2.3 \times 10^{8}$\,M$_{\odot}$ & (6) \\
M$_{\rm dust}$(60 and 100\,$\mu$m)& $2.5 \times 10^{6}$\,M$_{\odot}$ & (7) \\
L$_{\rm FIR}$  & $1.3 \times 10^{9}$\,L$_{\odot}$ & (7) \\
$\alpha_{\rm J2000}$$^{\mathrm{d}}$ & 03$^h$42$^m$01.49$^s$ & New centre \\
$\delta_{\rm J2000}$$^{\mathrm{d}}$ &-47$^{\circ}$13$^{\prime}$20.2\arcsec\ & New centre \\
\hline
\end{tabular}
\begin{list}{}{}
\item[$^{\mathrm{a}}$] ($\alpha_{\rm J2000}$, $\delta_{\rm J2000}$) is the
phase tracking centre of our $^{12}$CO interferometric observations
\item[$^{\mathrm{b}}$]
Luminosity and mass values extracted from the literature
have been scaled to the distance of $D$ = 9.9\,Mpc.
\item[$^{\mathrm{c}}$]
(1) NASA/IPAC Extragalactic Database (NED, http://nedwww.ipac.caltech.edu/);
(2) Veron-Cetty \& Veron (1986);
(3) Buta \etal (2001);
(4) HyperLeda;
(5) Ryder \etal (1996);
(6) Bajaja \etal (1995), reduced to the conversion factor 2.3 10$^{20}$cm$^{-2}$/(Kkm/s);
(7) {\it IRAS} Catalog.
\item[$^{\mathrm{d}}$]
New adopted centre, coinciding with the continuum peak.
\end{list}
\end{table}
\end{center}

%
\begin{center}
\begin{table}
      \caption[]{Main dynamical features in \ngc}
\label{tab:ring}
\begin{tabular}{lll}
\hline
Feature  & Radius & PA($^\circ$) \\
\hline
Nuclear bar & 9\arcsec\ (430~pc) & 31 \\
Nuclear ring &  9.5\arcsec\ (460~pc) & 31 \\
Primary bar&  83\arcsec\ (4~kpc) & 94 \\
Inner ring&  108\arcsec\ (5.2~kpc) & 95 \\
Outer ring&  190\arcsec\ (9.1~kpc) & 15 \\
\hline
\end{tabular}
\end{table}
\end{center}

\subsection{NGC 1433}
\label{sample}
The system
\ngc\ is a nearby active barred galaxy; it is a member of the Dorado group 
which includes 26 galaxies (Kilborn \etal 2005). We selected it from 
a sample of low-luminosity AGN  spirals already detected in CO emission for its proximity and 
moderate inclination. It was classified as
Seyfert 2 by Veron-Cetty \& Veron (1986), because of its strong nuclear
emission lines and its high [NII]/H$\alpha$ ratio. However, Sosa-Brito \etal
(2001) prefered to classify it as LINER, because of its [OIII]/H$\beta$ ratio,
which is just at the limit between Seyfert and LINERs. 
 Liu \& Bregman (2005) detected the nuclear point source in X-rays with ROSAT.

The galaxy has a rich network of dusty filaments around the nucleus. 
Its morphology reveals conspicuous rings (Buta 1986, Buta \etal 2001); 
the presence of nuclear, inner, and outer rings is the motivation of its nickname of  the 
``Lord of Rings'' (Buta \& Combes 1996).  Table \ref{tab:ring} presents
the sizes and orientations of the main dynamical features. Near IR images have revealed a nuclear bar 
inside the nuclear ring, of radius $\sim$400~pc (Jungwiert \etal 1997).  
The ring is the site of a starburst and is patchy in UV (continuum HST
 image from Maoz \etal 1996). Thirty one compact sources contribute 12\% of the UV light. 
Inside the ring the dust traces a flocculent or multiple-arm nuclear spiral structure 
(HST image from Peeples \& Martini 2006). There is a peak of 6~mJy in radio continuum 
emission at 843~MHz in the centre (Harnett 1987), with a weak extension along the bar. 
The HI 21cm emission map (Ryder \etal 1996) reveals that the atomic gas 
is concentrated in the inner and outer rings, with some depletion in the nuclear 
ring and bar region. In contrast, the central region is filled with molecular 
hydrogen (Bajaja \etal 1995, CO SEST map). Our ALMA single pointing 
includes in its field of view (FOV) all the nuclear bar and nuclear spiral gas.

\section{Observations}
\label{obs}

The observations were carried out with the Atacama Large Millimeter/submillimeter Array
(ALMA) telescope in Cycle 0,
with 19 antennae, during June and July 2012. The galaxy
\ngc\ was observed simultaneously in CO(3-2), HCO$^+$(4-3), 
HCN(4-3), and continuum, with Band 7. The sky frequencies were
344.56~GHz, 355.46~GHz, 353.24~GHz, and 343.27~GHz, respectively. The observations
were done in three blocks, with a total duration of two hours.
For each period, \ngc\ was observed for 27~minutes; the median
system temperatures were T$_{sys}$ = 140, 230, and 160~K.

The observations were centred on the nucleus, with a single pointing covering a FOV
of 18\arcsec. The Cycle 0 extended configuration provides
in Band 7 a beam of 0\farcs56$\times$0\farcs42, with a PA of 85$^\circ$. 
The galaxy was observed
in dual polarization mode with 1.875~GHz total bandwidth per baseband, and
a velocity resolution of 0.488~MHz $\sim$0.42~km/s. The spectra were then smoothed
to 4.88~MHz (4.24~km/s) to build channel maps.

This choice of correlator configuration, selected to simultaneously observe 
three lines,  provided  
a velocity range of 1600~km/s for each line, but did not centre the lines
(200km/s on one side and 1400 km/s on the other) which is adequate
for a nearly face-on galaxy, and 1800 MHz bandwidth in the continuum.
The total integration time provided an rms of 0.09 mJy/beam in the continuum,
and $\sim$3 mJy/beam in the line channel maps (corresponding to $\sim$170 mK,
at the obtained spatial resolution).
The flux calibration was done with the nearby quasar J0334-401, which is
regularly monitored at ALMA, and resulted in 10\% accuracy.

The data were calibrated, and cleaned using first a mask at the 50~mJy emission level,
and then the 30~mJy level. The final cube is 360x360 pixels with 0\farcs1 per pixel in the plane
of the sky, and has 60 channels of 4.24~km/s width.
The data were  calibrated, imaged, and cleaned with the CASA software (v3.3; McMullin \etal 2007), and the analysis
was then finalized with the GILDAS software (Guilloteau \& Lucas 2000).

The final maps were corrected for  primary beam attenuation to compute fluxes,
but were kept uncorrected for the plots. 
Almost no CO(3-2) emission was detected outside the full-width half-power (FWHP) primary beam. 
Because of  missing short spacings, extended emission
was filtered out at scales larger than $\sim$3\arcsec\,  in each 
channel map. The elongated features
corresponding to the dust lanes that were detected along the arms and rings are, however, quite narrow (thinner than 
2\arcsec\ as in HST images), so the 
missing-flux problem might not be severe in individual velocity slices.
Low-level negative sidelobes adjacent to bright emission were
observed.

\begin{table}
      \caption[]{CO(3-2) line fluxes, after primary beam correction}
\label{tab:line}
\begin{tabular}{lcccc}
\hline
\scriptsize{Line}& S$_{CO}$ &  V$_{hel}$ & $\Delta$V$^{(1)}$ & Peak flux \\
     &  Jy km/s &   km/s &   km/s  & Jy \\
\hline
Total &  234$\pm$ 1  & 1073.1$\pm$0.3 &85.3$\pm$ 0.7 &2.58 \\
\hline
 C1       &  103$\pm$ 2  & 1040.0$\pm$0.4   & 46.$\pm$ 1 &2.1 \\
 C2       &  105$\pm$ 4  & 1089.1$\pm$0.2   & 30.$\pm$ 1 &3.3 \\
 C3       &   26$\pm$ 3  & 1123.0$\pm$4.0   & 59.$\pm$ 5 &0.4 \\
\hline
 Blue$^{(2)}$   &  6.0$\pm$ 0.1  & 1018.7$\pm$0.6   & 61.$\pm$ 1 &0.09 \\
 Red$^{(2)}$   & 10.1$\pm$ 0.1  & 1138.2$\pm$0.3   & 56.$\pm$ 0.7 &0.17 \\
\hline
\end{tabular}
\begin{list}{}{}
\item[]   Total = Gaussian fit, assuming only one component,
C1/C1/C3 represent three velocity-component decomposition
\item[] $^{(1)}$ Full Width at Half Maximum FWHM 
\item[] $^{(2)}$ Fits for the blue and red components of the outflow, summed
over a region 0\farcs7$\times$1\farcs2 each (cf Fig. \ref{fig:spectrum}).
\end{list}
\end{table}

\begin{figure}[h!]
\centerline{
\begin{tabular}{c}
\includegraphics[angle=-90,width=8cm]{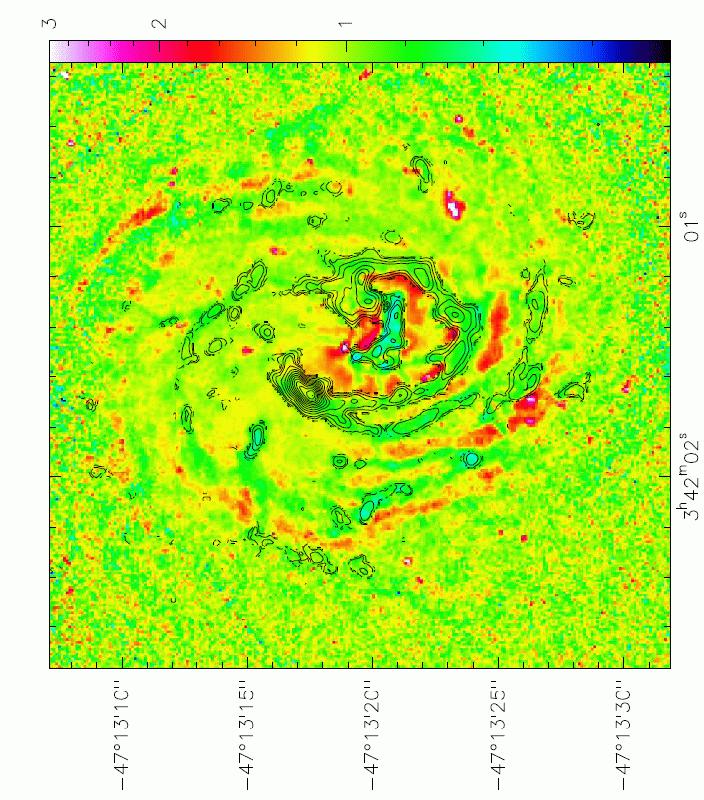}\\
\includegraphics[angle=0,width=8cm]{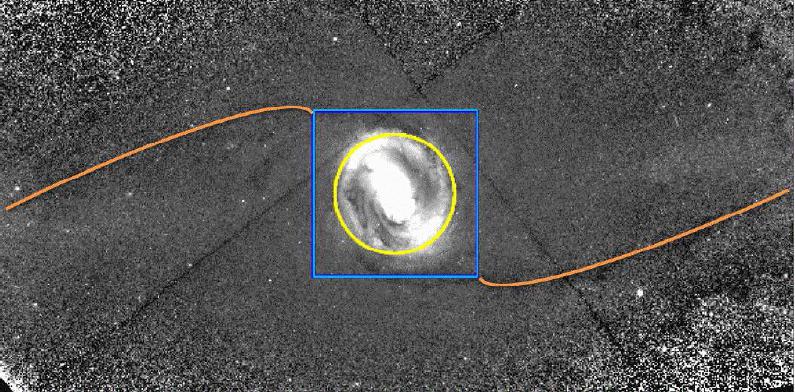}\\
\end{tabular}
}
\caption{{\it Top}: Overlay of CO(3-2) contours on the unsharp-masked blue (F450W) HST image.
The HST image has been aligned to correspond to the ALMA astrometry.
{\it Bottom:} Unsharp masking of the HST I-image of \ngc, covering the nuclear ring
and the dust lanes along the primary bar. 
The FWHP of the primary beam is indicated in yellow (18\arcsec\ in diameter), and the 
FOV of the CO map in Fig. \ref{fig:velo} and in the above image 
is indicated in blue (square of 24\farcs8 on a side).
The characteristic dust lanes on the leading edge of the main bar are outlined in orange. 
}
\label{fig:CO-on-B}
\end{figure}

\begin{figure}[h!]
\centerline{
\includegraphics[angle=0,width=8cm]{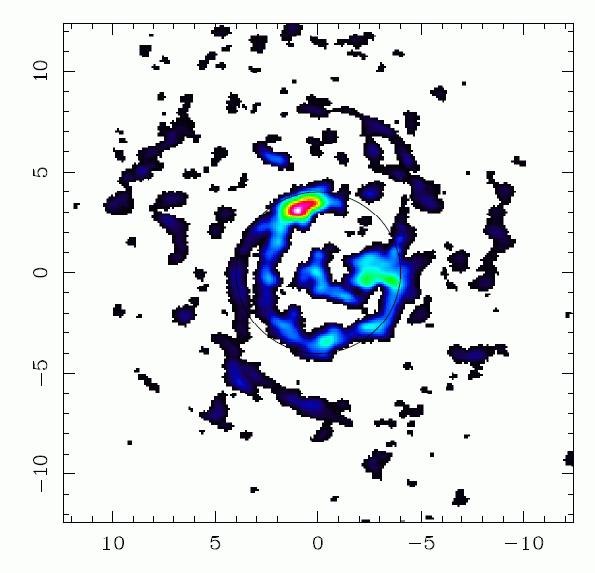}
}
\caption{Deprojection of the CO emission towards a face-on disk, centred on the new
adopted centre given in Table \ref{tab:basic}.
The pseudo-ring of 4\arcsec\ radius, here highlighted with a black circle, is standing out, nearly round.
}
\label{fig:deproj}
\end{figure}

\begin{figure}[h!]
\centerline{
\includegraphics[angle=-90,width=8cm]{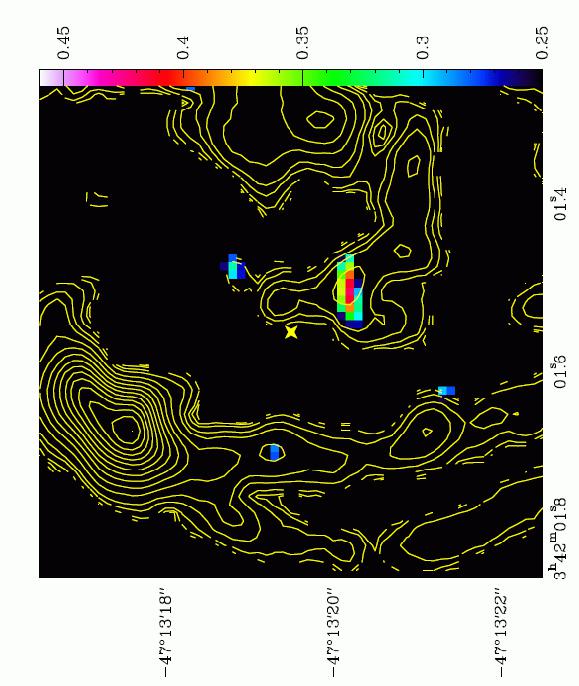}
}
\caption{Overlay of CO(3-2) contours on the 0.87mm continuum image. The FOV is
6\arcsec\ in diameter. The yellow star shows the phase centre given in Table \ref{tab:basic},
while the peak of the continuum is our new adopted centre. The colour palette 
unit is mJy.
}
\label{fig:cont}
\end{figure}

\section{Results}
\label{res}

Figure \ref{fig:chann} displays 42 of the CO(3-2) channel maps, with a velocity range
of  175~km/s and a velocity resolution of 4.24~km/s. The velocity field is 
regular, although perturbed by the tightly-wound spiral structure (see also Fig. \ref{fig:velo}).
At the outermost channels, the emission at the highest velocities does not occur primarily at large
radii, but mainly towards the centre.

\subsection{Molecular gas distribution and morphology}
\label{morpho}

To measure fluxes we used a clipped cube
where all pixel values $<$2$\sigma$ (6 mJy/beam) were set to zero. 
The mean intensity is plotted in Fig. \ref{fig:velo} (bottom).
Since the galaxy is more extended than the primary beam, it is difficult 
to quantify the missing flux. We compare it to the central spectrum
obtained with a single dish in Sect. \ref{CO-mass}; however, these
observations were obtained  with the SEST in CO(1-0) with a 43\arcsec\, beam.
Nevertheless, our FOV encompasses
the entire nuclear ring, and the emission in this nuclear region has by far
the strongest surface density at many wavelengths (Buta \etal 2001, Comeron \etal 2010;
Ho \etal 2011). 

We superposed the CO map onto the HST maps in the B, V, and I filters.
All show a remarkable similarity in morphology, as displayed in Fig. \ref{fig:CO-on-B}.
The features are so distinct that they were used to align the HST images which suffered from an
inexact astrometry.  The CO emission nicely corresponds to the dust lanes,
interleaved with the bright regions. The gas and the dust are closely mixed,
and reveal a multi-arm structure with a low pitch angle. There is not a well-defined density-wave here,
but rather a more flocculent spiral structure with multiple branches.  The structure is 
easily appreciated, thanks to the low inclination of the galaxy (33$^\circ$, Table \ref{tab:basic}).
This spiral structure is entirely included inside the nuclear ring of $\sim$ 10\arcsec=0.5 kpc in radius
(Buta \etal 2001, see Table \ref{tab:ring}).

What is remarkable is the large difference between the gas complex morphology in this nuclear region,
revealed by ALMA and the already known smoother stellar morphology (Buta \etal 2001). The lower panel of Fig. \ref{fig:CO-on-B}
shows an unsharp-masked red image of the nuclear region, embedded in the primary bar
whose leading dust lanes are marked. As is frequently found in strong primary
bars of early-type spirals, the dust lanes wind up onto the nuclear ring, which corresponds to the
inner Lindblad resonance, and a secondary nuclear bar decouples inside (e.g. Buta \& Combes 1996).
However, the gas does not follow the stellar nuclear ring, but instead flows through
a flocculent spiral onto an even smaller nuclear ring of $\sim$200pc radius, and from there reaches 
the very centre, at least at the 20pc scale, our resolution.

The molecular gas morphology reveals notable asymmetries; for instance, the peak of
CO emission is not in the centre but in a NE cloud complex, at about 4\arcsec\ from the centre
(200~pc), with no SW counterpart. In the very centre, the emission extends 2\arcsec\
to the SW, but with a corresponding hole in the NE. This might indicate an $m=1$ Fourier 
component in addition to the  $m=2$ and $m=3$ arm features.
To determine whether one particular $m$ component dominates, we have computed 
the Fourier decomposition of the 2D gas density once the galaxy disk has been
deprojected to the sky plane
\footnote{The decomposition is performed using the new centre defined in Sect. \ref{continuum}.}.
Figure \ref{fig:deproj} displays
the face-on molecular gas distribution. The pseudo-ring at radius $\sim$4\arcsec\, 
corresponding to 200~pc is clearly visible and nearly round. 
We have computed the radial distribution
of the various Fourier components, normalized to the axi-symmetric power.
The surface density of the gas has been decomposed as
$$
\Sigma(r,\phi) = \Sigma_0(r) + \sum_m a_m(r) cos(m\phi-\phi_m(r))
$$
and the amplitude of the various Fourier components $m$ are normalized as 
$A_m(r) = a_m(r)/\Sigma_0(r)$.  As a result, all $A_m(r)$ coefficients show
noisy behaviour, at a maxium amplitude of 0.5, but
there is no particular dominance of any $m$ feature.

\subsection{Continuum emission}
\label{continuum}

Besides the CO(3-2) line, continuum emission was detected at 0.87mm.
For that, the fourth band of width 468.8 MHz was used, with a rms noise level
of 0.15 mJy.
Figure \ref{fig:cont} displays the CO(3-2) contours superposed onto the
continuum map.  The peak emission is just detected at 3$\sigma$, about 0.5 mJy.
 The emission is extended in the east-west direction; its size is 1\arcsec x 0\farcs5.

\subsubsection{Recentring}
To establish the origin of the continuum emission, one issue is to determine
the exact position of the AGN.  We observed with a phase centre corresponding to the peak
of the near-infrared emission of the stellar component, which is known only
within 0\farcs7 uncertainty (e.g. 2MASS catalog, 2003). The HST maps, in B, V, and I would be precise
enough, but they are all affected by dust obscuration. In particular there is a 
conspicuous dust-lane extending nearly horizontally in the SW.
The continuum emission peaks at a position (-0\farcs6, -0\farcs7) with respect to our
phase centre, so perfectly compatible within the uncertainty.  This position is, however,
better centred with respect to the CO emission. We therefore
choose to adopt the peak of the continuum emission as the new centre.
 This is also perfectly compatible with the position of the X-ray nuclear point
source seen by Liu \& Bregman (2005).
Although the AGN might not correspond exactly to the peak of the stellar component, it is possible
that our new centre is also the correct position of the AGN and the supermassive black hole.
However it is unclear whether or not the 0.87\,mm AGN synchrotron emission is detected.

\subsubsection{Slope of radio-continuum emission}
Radio continuum emission has been detected at 35\,cm by Harnett (1987) with a resolution of 43\arcsec$\times$58\arcsec;
the emission is extended, and shows 6 mJy in the central beam. The galaxy
\ngc\ has also been observed at 21\,cm with ATCA by Ryder \etal (1996),
with a spatial resolution of 30\arcsec. The central emission is 3.4 mJy, quite similar to 
what is obtained at the ends of the bar from the HII regions. Since the whole nuclear region
is included in their central beam, it is possible that all the radio emission comes from 
star formation in the ring or nuclear region (both synchrotron from supernovae, and free-free emission). 
The continuum becomes 2 mJy at 4.8 GHz, with no polarisation (Stil \etal 2009). Comparing the
central fluxes at 21cm and 0.87mm, the slope of the radio spectrum would be -0.35, which could be a mixture
of synchrotron with a steeper spectrum ($-0.7$), and free-free emission with slope $-0.1$. 
Both steep radio spectra  (Sadler \etal 1995), and flat ones (Ulvestad \& Ho 2001)
have been found in Seyfert spiral galaxies, so it is not possible to conclude
on the AGN contribution in the centre.
From the H$\alpha$ flux it would be possible in principle to estimate
the fraction of free-free emission expected in the centre, but the spatial resolution (2\arcsec) is not enough
to disentangle what actually comes from the very centre. The extinction might also be a problem.

\subsubsection{Dust continuum emission}
Another possibility is that the continuum comes from thermal dust emission.
 At millimeter wavelengths we are nearly in the Rayleigh-Jeans domain, and the 
dust emission is only proportional to the dust temperature.  Continuum dust 
emission is then expected to be quite similar in morphology to the CO(3-2) emission 
(e.g. Dumke \etal 1997).   Why is this not the case?  The difference might be due to the lack of short spacing data,
and the filtering out of the diffuse extended continuum emission. Indeed, the continuum
is much more sensitive to this problem than the line emission.
From the IRAS fluxes, the average temperature of the dust in \ngc\ can be
estimated as 24~K, assuming that the dust opacity has a dependence in frequency
of $\nu^\beta$, with $\beta=2$. 
This is similar to central dust temperatures observed in $\sim$40\arcsec\ beams
with {\it Herschel} in star-forming barred galaxies such as
NGC\,3627 (Hunt \etal 2013, in prep.).
From a flux of 0.5 mJy/beam, and assuming the same
Draine \& Lee (1984) dust absorption cross section as described in Dumke \etal (1997)
for a solar metallicity, we find a molecular gas column density of 
N(\hh) = 4.5 10$^{22}$ cm$^{-2}$, over a beam of 24~ pc in size. This is what
is expected from a typical giant molecular cloud. In comparison, in the same 
position, the CO(3-2) emission is about 4 Jy\,km/s, for a CO integrated intensity in one
beam of 262 K\,km/s, corresponding to N(\hh) = 6 10$^{22}$ cm$^{-2}$, with 
a standard conversion factor of 2.3 10$^{20}$ cm$^{-2}$/(K\,km/s) (e.g.
Solomon \& Vanden Bout 2005). 
Considering all the uncertainties, the continuum emission is at the level 
expected from dust alone. Given that dust emission is only detected at the very centre,
it might be possible that this dust is associated with the molecular torus expected to hide the AGN
in this Seyfert 2 galaxy.  The derived mass of the torus would
be 9 10$^5$ M$_\odot$. Since the dust in the torus is certainly warmer than in the disk,
this might also explain why the continuum emission is not more extended, as is the CO, in addition
to the interferometer's filtering argument explained above. Mid-infrared maps with ISO at 7 and 15$\mu$m
also show a high central concentration, but with low resolution (Roussel \etal 2001).  

Only high-resolution observations with ALMA at several different frequencies would be able  
to settle the origin of the continuum emission and determine whether the AGN is detected directly.

\subsection{CO kinematics: a molecular outflow?}
\label{COdet}

In a previous paper (Buta \etal 2001), a detailed mass model of \ngc\ was performed
from NIR photometry and H$\alpha$ spectroscopy. Rotational and epicyclic frequencies 
($\Omega$ and $\kappa$) were then derived, and together with the numerical simulations
from Buta \& Combes (2000), the predictions of the resonance locations, compared
to the observed ring radii, favoured a pattern speed of 23~km/s/kpc (or 26~km/s/kpc with our slightly different
distance adopted). With this pattern speed, there are two inner Lindblad resonances (ILRs)
located at 3.6 and 30\arcsec\footnote{Treuthardt \etal (2008) propose a lower value for the pattern speed in \ngc,
18~km/s/kpc, but their simulation shows a nuclear ring much larger in size than observed.}.
The existence of two ILRs weakens the primary bar, and allows the
decoupling of a secondary bar, with a higher pattern speed (e.g. Friedli
\& Martinet 1993, Buta \& Combes 1996). The nuclear bar produces negative torques
on the gas, previously stalled at the nuclear ring, and provides
a dynamical way to fuel the nucleus. This process has been simulated 
in detail in Hunt \etal (2008), and shows how the gas in the nuclear ring
progressively flows to the centre, in a spiral structure and in 
a ring shrinking in radius. It appears that this scenario applies quite well
to \ngc: its nuclear ring lies between the two ILRs,
and the molecular gas morphology reveals an accumulation of the gas at the inner ILR.
This configuration strongly suggests that the gas is presently fueling the AGN.

\begin{figure}[h!]
\centerline{
\begin{tabular}{c}
\includegraphics[angle=-90,width=8cm]{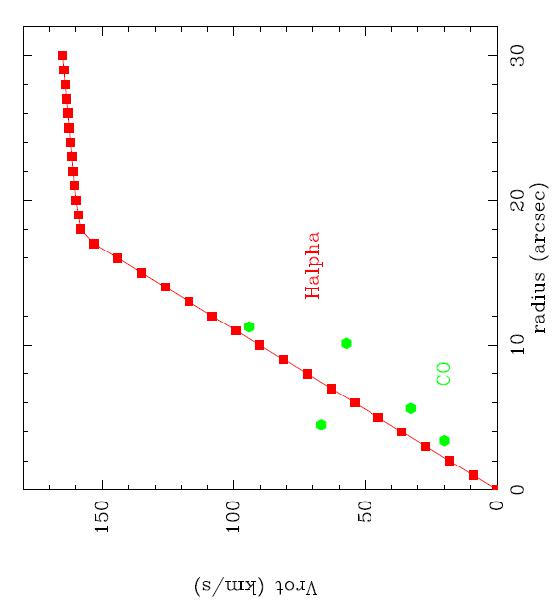}\\
\includegraphics[angle=-90,width=8cm]{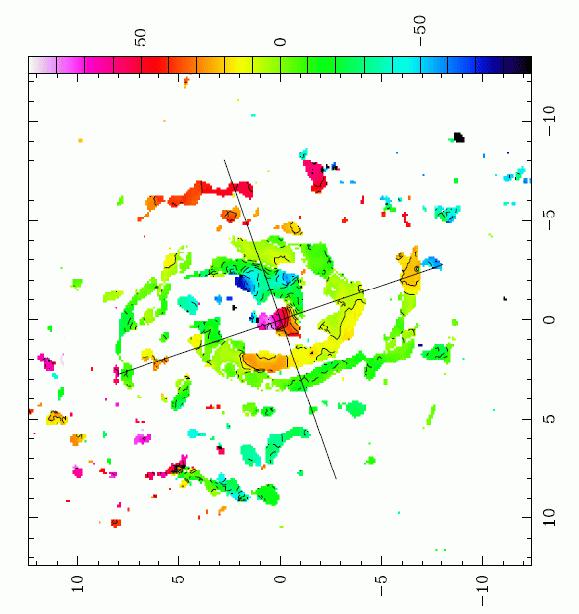}\\
\end{tabular}
}
\caption{{\it Top}: Rotational velocity model adopted for \ngc,
based on the H$\alpha$ kinematics (red filled squares) from Buta \etal (2001),
compatible with the CO rotation curve (green filled hexagons).
 The CO velocity field is, however, sparsely sampled.
{\it Bottom}: Velocity residuals after subtraction of a regular rotation model,
based on the H$\alpha$ rotation curve above. The map has been recentred
on the new adopted centre given in Table \ref{tab:basic}.  The two orthogonal lines
 indicate the position of the PV diagrams in Fig. \ref{fig:pv}.   
}
\label{fig:vrot}
\end{figure}

The top panel of Fig. \ref{fig:velo} displays the velocity field of
the molecular gas.  The velocity field is well described by rotation, with the same 
position angle as the HI velocity field at larger scales (Ryder \etal 1996) and is consistent with the H$\alpha$
central kinematics (Buta \etal 2001). Superposed on this regular rotation, there are no strong perturbations 
due to streaming motions in a barred potential, since the major axis of the galaxy is aligned
with the minor axis of the primary bar, and also with the nuclear bar
(see Fig. \ref{fig:CO-on-B}).
The amplitude of the rotation is low but compatible with the observed H$\alpha$ velocities
within 10\arcsec\ in radius, given the low inclination of 33$^\circ$. 
For purposes of comparison, the rotation
velocities deduced from the CO kinematics and the
H$\alpha$ rotation curve are shown together in Fig. \ref{fig:vrot}.

\begin{figure}[h!]
\centerline{
\begin{tabular}{c}
\includegraphics[angle=0,width=8cm]{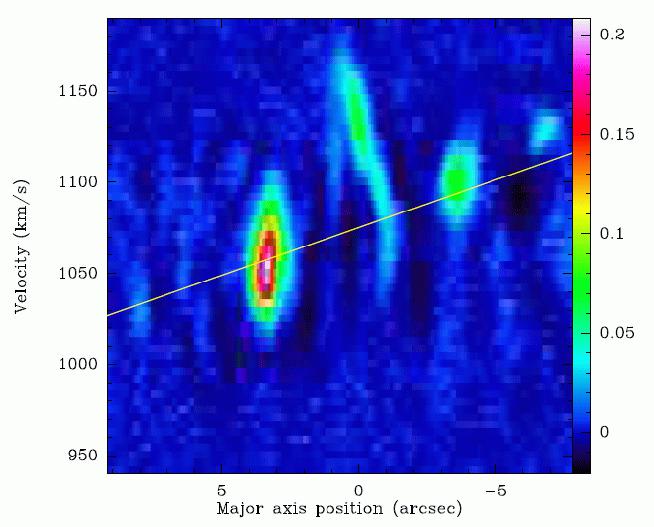}\\
\includegraphics[angle=0,width=8cm]{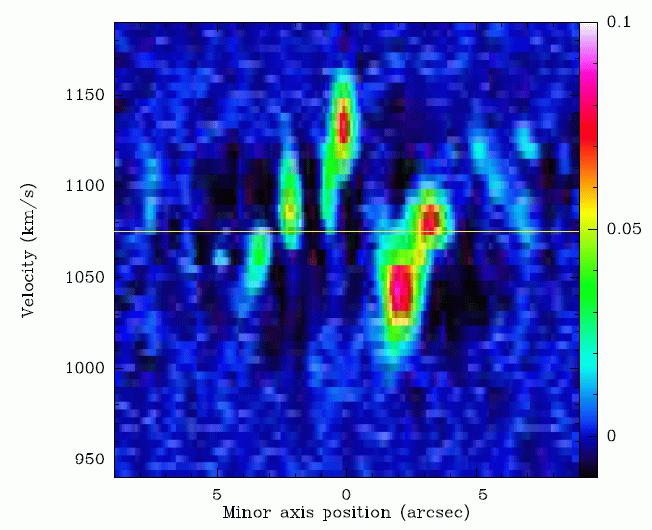}\\
\end{tabular}
}
\caption{{\it Top}: Position-velocity diagram along the major axis of PA = 199$^\circ$
(east is at left).
The central outflow is clearly visible superposed on the smooth rotational velocity gradient
(underlined by the yellow line, corresponding to the rotation curve of Fig. \ref{fig:vrot}).
{\it Bottom}: Position-velocity diagram, along the minor axis, of PA = 109$^\circ$
(east is at left). The two components (red and blue) of the outflow are visible,
along a slice, where the velocity should be equal to the systemic one (yellow line).
}
\label{fig:pv}
\end{figure}

There is, however, a noticeable redshifted perturbation located at the very centre 
and it extends to the south-west
between 0 and 2\arcsec\, i.e. 100~pc in extent. To better isolate this feature, we plot
the position-velocity diagram along the major axis of the galaxy in the top panel of
Fig. \ref{fig:pv}.
This slice reveals gas jumping by almost 100~km/s in projection, at much higher velocity than the
rest of the nuclear disk gas. 
There is also a noticeable blue-shifted counterpart, at a distance from the centre of about 2\arcsec\ (100~pc)
towards the north-west,
which is conspicuous in the position-velocity diagram along the minor-axis of the galaxy in the bottom panel of
Fig. \ref{fig:pv}. Both flow components are seen in this direction, while the largest
gradient of velocities is along PA= 135$^\circ$, which might be the projected direction of the flow.

Another way to compare these peculiar velocities to the rest of the nuclear region in 2D
is to subtract the expected regular velocity field known from the H$\alpha$ gas in the
same region.  Figure \ref{fig:vrot} displays the residuals obtained, relative to the adopted
H$\alpha$ rotation curve, plotted above.  The figure shows the
ionized gas rotation curve deduced by Buta \etal (2001).
 The derived CO velocities, although in sparse regions, 
are compatible with this adopted rotation curve.  The stellar velocity, 
once corrected for a large asymmetric drift, appears higher (Buta \etal 2001).
The gas then does not follow the maximum
circular velocity. This might be due to substantial gas turbulence, 
and/or to an overestimation of the correction of the stellar velocity.

The peculiar velocity of the gas at the nucleus and north-west of the centre is clearly seen in the residuals of
Fig. \ref{fig:vrot}.  If the gas is in the plane, the deprojected velocity could
be as high as 200~km/s, but other orientations with respect to the sky plane are possible.
We call $\alpha$ the angle between the outflow direction and the line of sight.
The observed velocity in projection is V$_{outflow}$cos($\alpha$), and 
the extent of the flow in the plane of the sky is R$_{outflow}$sin($\alpha$).
 It is likely that $\alpha$ is not close to the
extreme values, i.e. zero or 90 degrees, since the observed outflow velocity
and the projected size of the outflow are both subtantial, i.e.  $\sim$ 100~km/s
and $\sim$ 100~pc, respectively. This means that tan($\alpha$) is of the order of 1.
The flow is aligned roughly with the minor axis, and if it is orthogonal to the plane,
tan($\alpha$)=0.6. We think, however, that the outflow is not orthogonal,
since we are seing the galaxy inclined by 33$^\circ$  on the sky, and the near side is the NW,
from the winding sense of the spiral arms, assumed trailing.
The outflow cannot be exactly perpendicular to the disk, unless the blue
and red regions are inverted. The flow must at least
be inclined by an angle $>$ 33$^\circ$ from the normal to the plane. 
 Conservatively, the outflow velocity probably lies  between 100~km/s and 200~km/s. 

\begin{figure}[h!]
\centerline{
\includegraphics[angle=0,width=8cm]{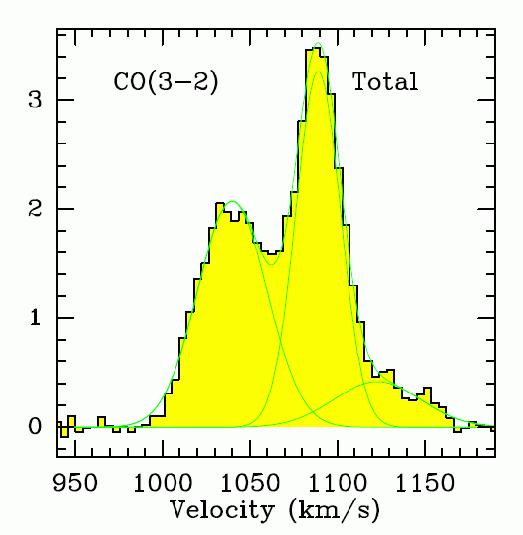}
}
\caption{Total CO(3-2) spectrum, integrated over the observed map,
with a FOV of 18\arcsec, after correction for primary beam attenuation. 
The vertical scale is in Jy. The green line is the result
of the Gaussian fit with three velocity components; see Table \ref{tab:line}.
}
\label{fig:spectot}
\end{figure}

This high-velocity gas is also noticeable in the total spectrum, obtained
by summing the signal over the field of view, as in Fig. \ref{fig:spectot}.
A Gaussian decomposition in three components has been performed on the 
spectrum, and the results are displayed in Table \ref{tab:line}.  The high-velocity red 
component represents nearly 5\% of the total. The blue-velocity counterpart is
diluted in the normal rotational component C1 (part of the two-horn profile characteristic of rotation).

\begin{figure*}[ht!]
\centerline{
\includegraphics[angle=-0,width=15cm]{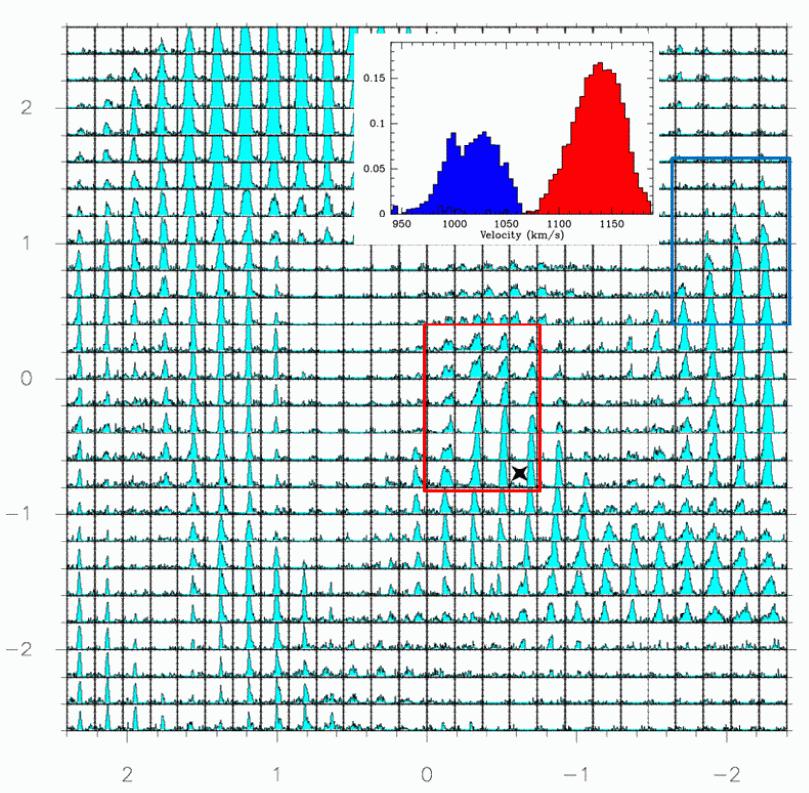}
}
\caption{CO(3-2) spectra within 2\farcs5 of the centre. The velocity scale
is from 960 to 1190~km/s (corresponding to a range from -115 to 115~km/s with respect to the systemic
velocity).  The vertical scale is from 1 to 40 mJy. The new adopted centre is marked
with a black star (the map coordinates are with respect to the phase centre).
 The red velocity component can be seen at the new centre and just above,
and the blue velocity component is centred at (-2, 0.5). The two regions of 
0\farcs7$\times$1\farcs2 each, selected to integrate the outflow mass in Table \ref{tab:line}, are
indicated by red and blue rectangles, respectively. The corresponding integrated spectra
are plotted in the insert (scale in Jy).
}
\label{fig:spectrum}
\end{figure*}

Finally, the high-velocity red component is best located in the map through
the individual spectra of Fig. \ref{fig:spectrum}, where all spectra are
shown within a radius of 2\farcs5.
Although in principle, this high-velocity gas could be inflowing as well as outflowing, we
consider inflow to be unlikely. Indeed, there is no other signature of violent perturbation
due to a companion nearby, and if gas were slowly accreted to fuel the AGN, it would
first have settled into the rotational frame at much larger radii than the last 100~pc.

To better estimate the quantity of gas in the outflow, we have summed the
CO flux within two regions of sizes
0\farcs7$\times$1\farcs2, centred  on the red and  blue outflow regions, taking
into account the primary beam correction (cf Fig. \ref{fig:spectrum}). The results are given
in Table \ref{tab:line}. 
Assuming the standard CO-to-\hh\ conversion factor (see next section), we derive
molecular masses of 1.3 10$^6$ and 2.3 10$^6$ M$\odot$ for the blue and red velocity
components, respectively.

Is the outflow also detected in the ionized gas?  There is no outflow detected in 
X-rays, but there is not enough spatial resolution to see it anyway. In H$\alpha$
maps and spectroscopy, it is hard to reach a conclusion,
even from the best velocity field obtained from Fabry-Perot interferometry
by Buta (1986). In his Fig. 8, we can see a quite perturbed velocity field inside the central 
20\arcsec, which may reflect steep gradients. However, the spatial resolution is only 2\arcsec,
while the projected distance between our red and blue outflow peak components is roughly 
the same.  An outflow of ionized gas is, however, compatible with the data.
The non-detection of ionized gas outflow in galaxies showing a molecular outflow
is also found in other compact systems like NGC~1377 (Aalto \etal 2012). A comparison
with other molecular outflows will be discussed in Sect. \ref{disc}.

\subsection{CO luminosity, \hh\, mass, and HCO$^+$/HCN upper limits}
\label{CO-mass}

Figure \ref{fig:spectot} displays the total CO(3-2) spectrum, integrated
over the entire observed map  after correction for primary beam attenuation. 
 When integrated over the line (FWHM=85.3~km/s), the 
integrated emission is 234$\pm$ 1 Jy\,km/s.  Towards the central position, Bajaja \etal (1995)
found a CO(1-0) spectrum peaking at T$_A^*$= 48mK, with  FWHM=168~km/s,
yielding a total integrated flux of 193 Jy\,km/s, in a beam of 43\arcsec. Unfortunately, 
no CO(2-1) spectra have been reported. We can, however, remark that the CO(1-0) flux corresponds to a larger
region than the one observed here (as witnessed by the broader linewidth), and our CO(3-2) flux is an upper limit of 
the expected CO(1-0) flux in a 18\arcsec\, beam, since at low $J$ the flux increases with the $J$ level. 
We can therefore safely conclude that the CO(3-2)/CO(1-0) ratio
must be significantly larger than 2 in flux density units: the CO gas is relatively excited,
 meaning that the average density is at least of the order of 10$^4$ cm$^{-3}$. 
Assuming a CO(3-2)/CO(1-0) flux ratio of $\sim$5, similar to that
observed in the star-forming nuclei of nearby galaxies (e.g. Matsushita \etal 2004,
Boone \etal 2011), the CO detected inside our primary beam, at the distance of 9.9 Mpc, 
corresponds to a molecular mass M(\hh) = 5.2 10$^7$ M$_\odot$, with the standard CO-to-\hh\ 
conversion factor of 2.3 10$^{20}$cm$^{-2}$/(Kkm/s).
By comparison, Bajaja \etal (1995) find in their central 43\arcsec\ beam a mass of  1.8 10$^8$ M$_\odot$,
covering an area 5.7 times larger.

As far as the CO outflow is concerned, the use of any CO-to-\hh\ conversion factor is
uncertain. We then try to get a strict minimum of the mass in using the hypothesis of
optically thin emission. Within this hypothesis, we can write the 
column density of CO in the upper state of the (3-2) transition as:
$$
N_{J=3} = 6.4 10^{13} I_{CO(3-2)},
$$
where $I_{CO(3-2)}$ is the integrated (3-2) emission in K.km/s. The total CO column density 
is then obtained, using the ratio
$$
N_{J=3}/N_{CO} = \frac{5}{Q} exp(-E_{J=3}/kT_{ex}),
$$
where $E_{J=3}$ is the energy in the upper level of the (3-2) transition,
Q the partition function = 0.36 $T_{ex}$, and  $T_{ex}$ the excitation temperature
assumed constant over all $J$ levels.  The N(\hh) column density is then derived, assuming a CO abundance of 6 10$^{-5}$ (e.g.  Glover \& Mac Low 2011).
  Comparing the N(\hh) values obtained within the optically thick hypothesis, and the
use of the standard conversion factor, we found column densities lower by factors of
27, 71, and 83 when $T_{ex}$= 10, 20, and 30K, respectively. Over the red outflow region
that is as large as 3-4 beams, we found  N(\hh) $\sim$ 5 10$^{22}$ cm$^{-2}$, while the 
optically thin hypothesis will lead to values as low as  6 10$^{20}$ cm$^{-2}$.
We estimate that values this low are not realistic, however, since the mean 
volumetric density over the region will be $\sim$ 1 cm$^{-3}$, and not
100 cm$^{-3}$, the minimum required to excite CO emission. We note that the 
mean CO(3-2) brightness temperature observed within the flow region if 3K, so that
the surface filling factor of the molecular component cannot be much smaller than 0.1. Since the critical density of the CO(3-2) line is 10$^5$ cm$^{-3}$, 
the optically thin hypothesis is very unlikely to provide any emission,
even taking into account the surface filling factor.  

Finally, our simultaneous observations of HCO$^+$(4-3) and HCN(4-3) yielded only negative
results. We can better derive significant upper limits towards the CO emission maxima.
Over the whole map, there were 160 pixels (equivalent to 8 beams) with CO(3-2) emission larger than 60 times
the 3$\sigma$ upper limits in HCO$^+$(4-3) and HCN(4-3), assuming the same linewidth.
In all CO maxima, an intensity ratio between CO and the high-density tracers $>$60 means that the
average density of the gas in the multiple-arm flocculent spiral is not high. The critical density to excite
the HCO$^+$(4-3) and HCN(4-3) molecular lines is at least 10$^7$ cm$^{-3}$.

\section{Discussion and summary}
\label{disc}

We have presented our first ALMA results for a Seyfert 2 galaxy from our extended NUGA sample, \ngc.
The observations in CO(3-2) allow us to reach an unprecedented spatial resolution of 24 pc, even
with the limited Cycle 0 capabilities.

The morphology of the CO emission comes as a surprise. Although the Seyfert 2-type would suggest 
the presence of a thick obscuring component in front of the nucleus,
there is no large concentration of molecular gas in the centre, but instead a widely distributed
multiple-arm spiral of CO emission all over the nuclear ring region. The dense gas tracers 
HCO$^+$ and HCN remain undetected, confirming the absence of very dense gas (density 
larger than 10$^7$ cm$^{-3}$).

Although infrared images reveal the presence of a stellar nuclear bar inside the nuclear ring (of radius 0.5~kpc),
located near the inner Lindblad resonances (e.g. Buta \etal 2001), the gas does not follow the nuclear bar.
Instead the gas appears to flow inward and to partly accumulate in a ring-like structure at a radius $\sim$ 200~pc,
which coincides with the inner ILR (IILR) as computed by Buta \etal (2001). This is indeed expected
at some epochs of self-consistent N-body+hydro simulations, when the gas enters an inflowing
phase inside two ILRs (e.g. Hunt \etal 2008). The gas is not stalled in this pseudo-ring, but continues
to flow in towards the very centre.

The kinematics of the CO emission are dominated by a regular rotational velocity field,
with only slight perturbations from the multiple-arm spiral. No strong streaming motion is imprinted
on these kinematics by the primary and nuclear bars, since their axes coincide with the galaxy
major axis. Additionally, two peculiar features appear at high velocity, a red-shifted component
 towards the centre within 100~pc, and a blue-shifted counterpart at 2\arcsec\ (100~pc) from the centre. The
amplitude of these components is up to
about 100~km/s in projection ($\sim$ 200~km/s if in the galaxy plane). 
 Given their location near the nucleus,
we tentatively interpret these high-velocity features as the two sides of an outflow. Globally, these features 
represent as much as $\sim$ 7\% of the total molecular emission in the nuclear ring region,
i.e., 3.6 x 10$^6$ M$_\odot$.

It is not likely that these peculiar high-velocity features reflect strong streaming motions due to a
dynamical perturbation, since there is no perturbation of this kind in the centre.
The gas does not follow the nuclear bar, which is rather weak.   Is a central mass
able to generate such a high rotation in the centre? Considering that the blue and red components
are separated in projection along the minor axis by 2\arcsec = 100~pc,
or about 120~pc in the plane of the galaxy, a massive black hole located in the centre, at R=60pc from each component,
should have a mass of at least M$_{BH}$=V$^2$ R/G for the rotational velocity in the galaxy
plane V=200km/s, or M$_{BH}$=5.6 10$^8$ M$_\odot$.
This would make \ngc\ a strong outlier to the M$_{BH}-\sigma$ relation; indeed
from the bulge mass, we would expect the BH mass to be 5 10$^6$ M$_\odot$, 
e.g., Buta (1986). In any case, for gas rotating in circular motion
within the sphere of influence of the black hole, the velocity maxima should appear on the
major axis and disappear on the minor axis, contrary to what is observed here. 
Another solution would be to assume the existence of a mini-polar disk with completely 
different orientation to the main disk, and almost edge-on, but no galaxy interaction or
accretion event supports this hypothesis.

The origin of the outflow might  be related to star formation, which is concentrated in
the nuclear ring region. The star formation rate (SFR) can be estimated from the far infrared luminosity
as calibrated by Kennicutt (1998). From the IRAS fluxes, the FIR luminosity is
1.3 10$^9$ L$_\odot$ (Table \ref{tab:basic}), and the SFR equals 0.2  M$_\odot$/yr.
From the H$\alpha$ luminosity, measured at 3.7 10$^{40}$ erg/s by Hameed \& Devereux (2005),
we can also deduce from Kennicutt's calibration a SFR = 0.29 M$_\odot$/yr,
which is compatible.

The order of magnitude of the mass outflow rate can be computed, using our estimates for the molecular
mass in the high-velocity components (Table \ref{tab:line}), as M=3.6 10$^6$ M$_\odot$. 
This mass was obtained using the standard CO-to-\hh\, conversion factor, since there is
 no reason a priori to adopt the lower factor applying to ultra-luminous galaxies. Cicone \etal (2012) show
in Mrk231 that the molecular gas of
the galaxy and the outflowing gas share the same excitation.
However, this mass
could be an upper limit if the flow is made of more diffuse gas. 
Since each high-velocity component has a projected radial extent from the 
centre of d=1\arcsec\ $\sim$ 50~ pc, and moves at a projected velocity
of v=100~km/s, the flow rate is of the order of dM/dt $\sim$ (Mv/d) tan$\alpha$= 7 tan$\alpha$ M$_\odot$/yr,
with $\alpha$ being the angle between the outflow and the line of sight. 

 Although
this estimate is uncertain by a factor of a few, given the unknown $\alpha$,
it is about 40 times higher than the SFR; since galactic winds due to starbursts correspond in
general to mass outflows of the same order as the SFR (e.g. Veilleux \etal 2005), we conclude
that the outflow is probably not due to star formation alone, and is at least helped by the AGN.
We note that starburst winds are generally observed in galaxies
with SFR larger than  5 M$_\odot$/yr, and SFR surface densities larger
than 10$^{-3}$  M$_\odot$/yr/kpc$^2$. The galaxy \ngc\, has a low total SFR  $\sim$ 0.2  M$_\odot$/yr;
however, its SFR surface density is 0.34  M$_\odot$/yr/kpc$^2$ if we assume that the whole SFR is 
confined to the nuclear disk of 9\arcsec radius. The SFR surface density would therefore be
enough to drive a wind, although (as noted above) the mass loading
factor expected for this type of wind would still be considerably
lower than what we observe in NGC\,1433. 

The kinetic luminosity of the flow can be estimated as L$_{kin}$ =0.5 dM/dt v$^2$
=2.3  tan$\alpha$ (1+ tan$^2\alpha$) 10$^{40}$ erg/s.  The luminosity of the AGN 
can be estimated at various wavelengths. Although the X-ray point source is weak, 1.7 10$^{39}$ erg/s
over 0.3-8kev (Liu \& Bregman 2005), we can derive a bolometric luminosity of the 
AGN from  optical and NIR magnitudes in the central aperture (Buta \etal 2001) of  1.3 10$^{43}$ erg/s. From the expected
BH mass of 5 10$^6$ M$_\odot$, if \ngc\, is on the  M$_{BH}-\sigma$ relation, the Eddington luminosity is
6.3  10$^{44}$ erg/s. The kinetic luminosity of the outflow is low with respect to the bolometric luminosity
of the AGN, making it plausible that the latter is able to power the wind.

The momentum flux of the outflow, computed by  dM/dt v is larger compared
to that provided by the AGN photons L$_{\rm AGN}$/c, by a factor of 10 tan$\alpha$/cos$\alpha$.
Since the momentum can be boosted in case of energy-conserved wind by factors up to 50 (e.g. Faucher-Gigu\`ere
\& Quataert 2012), it is possible that the AGN contributes to drive the outflow only by its radiation
pressure.
Alternatively, it is possible that the outflow is driven
through the AGN radio jets. From the central 1.4 GHz power of 3.4mJy detected by Ryder \etal (1996),
we can estimate the jet power from the formula proposed by Birzan \etal (2008,
their Eq. 16): P$_{jet}$ = 2 10$^{42}$ erg/s. Since this power is about two orders of magnitudes higher
than the kinetic luminosity of the outflow, the jet is amply able to drive the flow, even with
low coupling. The jet interaction with the interstellar medium has been simulated by Wagner \etal (2012)
who show that the jet is able to drive a flow efficiently as soon as the Eddington ratio
of the jet P$_{jet}$/L$_{Edd}$ is larger than 10$^{-4}$. In NGC~1433, this ratio is about 3.2 10$^{-3}$.
  
The molecular outflow in NGC~1433 is one of only a few discovered recently occuring in low-star forming
galaxies  with relatively weak AGN where the flow might be driven by both the starburst and
the radio jets.
The LINER NGC 6764 has 4.3 10$^6$ M$_\odot$ of molecular gas driven out with a velocity of about
100~km/s (Leon \etal 2007). The flow projects to larger distances than in NGC~1433, and might be more
evolved. The outflow rate is lower, of the order of  1 M$_\odot$/yr. The galaxy NGC 1266 is also a LINER and has the highest flow
rate of 13  M$_\odot$/yr, with 2.4 10$^7$ M$_\odot$ of molecular gas driven with V=177 km/s
(Alatalo \etal 2011).
A third LINER with total SFR of $\sim$ 1 M$_\odot$/yr, NGC 1377 has an outflow rate of
8 M$_\odot$/yr, and an outflowing mass of 1.1 10$^7$ M$_\odot$ at V=140 km/s (Aalto \etal 2012).
All these galaxies have star formation playing a role in the outflow, but the properties
of the flow require the contribution of the AGN through the entrainement of its radio jets.
This is the most needed for NGC~1433, which has the lowest SFR of all.

This tentative detection of a molecular gas outflow triggered essentially by  the AGN, 
should be confirmed by higher-resolution ALMA observations. The detection of a radio
continuum component at the very centre, which might be due to thermal dust emission 
from a molecular torus, also deserves a higher resolution study.

\begin{acknowledgements}
  We warmly thank the referee for constructive comments and suggestions. 
The ALMA staff in Chile and ARC-people at IRAM are gratefully acknowledged for their
help in the data reduction. We particularly thank Gaelle Dumas and Philippe 
Salom\'e for their useful advice.
This paper makes use of the following ALMA data: ADS/JAO.ALMA\#2011.0.00208.S.
ALMA is a partnership of ESO (representing its member states), NSF (USA) and NINS (Japan),
together with NRC (Canada) and NSC and ASIAA (Taiwan), in cooperation with the Republic of
Chile. The Joint ALMA Observatory is operated by ESO, AUI/NRAO and NAOJ.
The National Radio Astronomy Observatory is a facility of the National Science Foundation
operated under cooperative agreement by Associated Universities, Inc.
We used observations made with the NASA/ESA Hubble Space Telescope, and obtained 
from the Hubble Legacy Archive, which is a collaboration between the Space Telescope 
Science Institute (STScI/NASA), the Space Telescope European Coordinating Facility 
(ST-ECF/ESA), and the Canadian Astronomy Data Centre (CADC/NRC/CSA). 
F.C. acknowledges the European Research Council
for the Advanced Grant Program Num 267399-Momentum.
I.M. acknowledges financial support from the Spanish grant AYA2010-15169
and from the Junta de Andalucia through TIC-114 and the Excellence Project
P08-TIC-03531.
We made use of the NASA/IPAC Extragalactic Database (NED),
and of the HyperLeda database.
\end{acknowledgements}

\end{document}